\documentstyle[prl,aps,multicol,epsfig]{revtex}
\begin{document} 
\draft 
\title{Paramagnetic metal-insulator transition in the 2D Hubbard model}

\author{C. Gr\"ober, M. G. Zacher, and R. Eder}
\address{Institut f\"ur Theoretische Physik, Universit\"at W\"urzburg,
Am Hubland,  97074 W\"urzburg, Germany}
\date{\today}
\maketitle

\begin{abstract}
We study the transition from paramagnetic metal to paramagnetic insulator 
by finite temperature Quantum Monte-Carlo simulations for the 2D Hubbard 
model at half-filling. Working at the moderately high temperature $T=0.33t$, 
where the spin correlation length has dropped to $< 1.5$ lattice 
spacings, and scanning the interaction strength $U$ 
we observe an unambiguous metal-insulator transition with onset at 
$U_c\approx 4t$. In the metallic phase there are no indications for any 
`correlation narrowing' of the band width, nor for any decrease of spectral 
weight at the Fermi energy. The Mott-Hubbard gap opens gradually with $U$ and 
is accompanied by the formation of side bands near the Fermi energy. 
\end{abstract}
\begin{figure}
\epsfxsize=6.0cm
\vspace{-0.0cm}
\hspace{-0.5cm}\epsfig{file=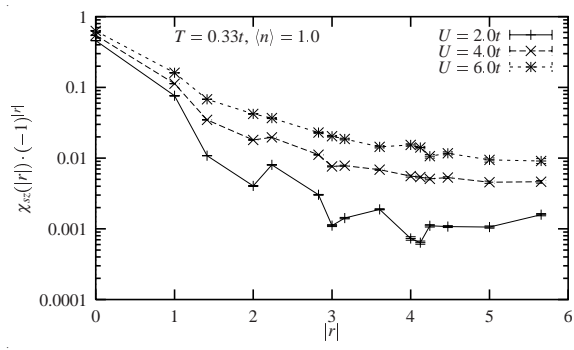,width=8.0cm,angle=0.0}
\vspace{0.5cm}
\narrowtext
\caption[]{Equal-time spin correlation function
$\chi_{sz}(\bbox{r})$ $=$ $(1/N)$ $\sum_i
\langle \bbox{S}_i \cdot \bbox{S}_{i+\bbox{r}}\rangle$
in the $8\times 8$ cluster for different $U/t$.}
\label{fig1} 
\end{figure}

\begin{figure}
\epsfxsize=6.0cm
\vspace{-0.0cm}
\hspace{-0.5cm}\epsfig{file=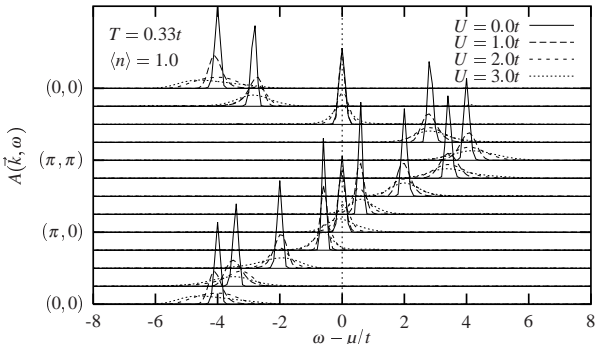,width=8.0cm,angle=0.0}
\vspace{0.0cm}
\narrowtext
\caption[]{Single particle spectral function $A(\bbox{k},\omega)$
Along high-symmetry directions of the Brillouin zone,
for the $8\times 8$ clusters and different values of $U/t$.}
\label{fig2} 
\end{figure}

\begin{figure}
\epsfxsize=6.0cm
\vspace{-0.0cm}
\hspace{-0.5cm}\epsfig{file=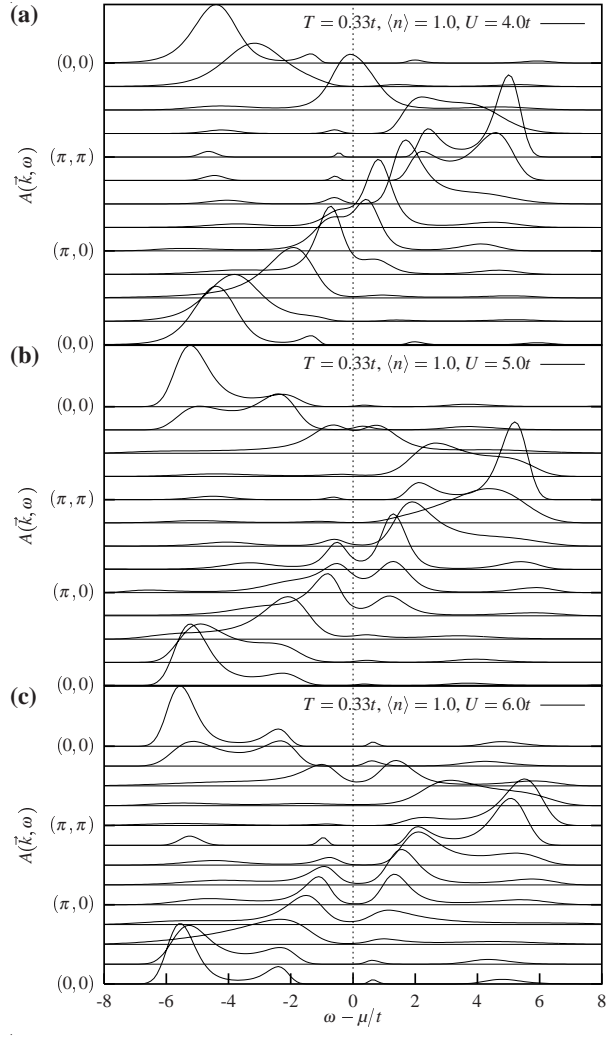,width=8.0cm,angle=0.0}
\vspace{0.0cm}
\narrowtext
\caption[]{Same as Figure \ref{fig2} but for larger values of $U/t$.}
\label{fig3} 
\end{figure}

\begin{figure}
\epsfxsize=6.0cm
\vspace{-0.5cm}
\hspace{0.5cm}\epsfig{file=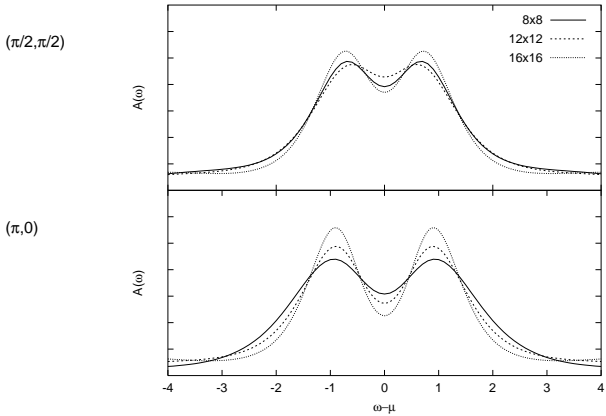,width=6.0cm,angle=270.0}
\vspace{0.0cm}
\narrowtext
\caption[]{Size dependence of the spectral function at $(\pi,0)$
and $(\pi/2,\pi/2)$ for $U/t$$=$$5$. For these two momenta particle
hole-symmetry implies $A(\omega)$$=$$A(-\omega)$, which was used as 
additional information in the MaxEnt reconstruction.}
\label{fig4} 
\end{figure}

\begin{figure}
\epsfxsize=6.0cm
\vspace{-0.0cm}
\hspace{-0.5cm}\epsfig{file=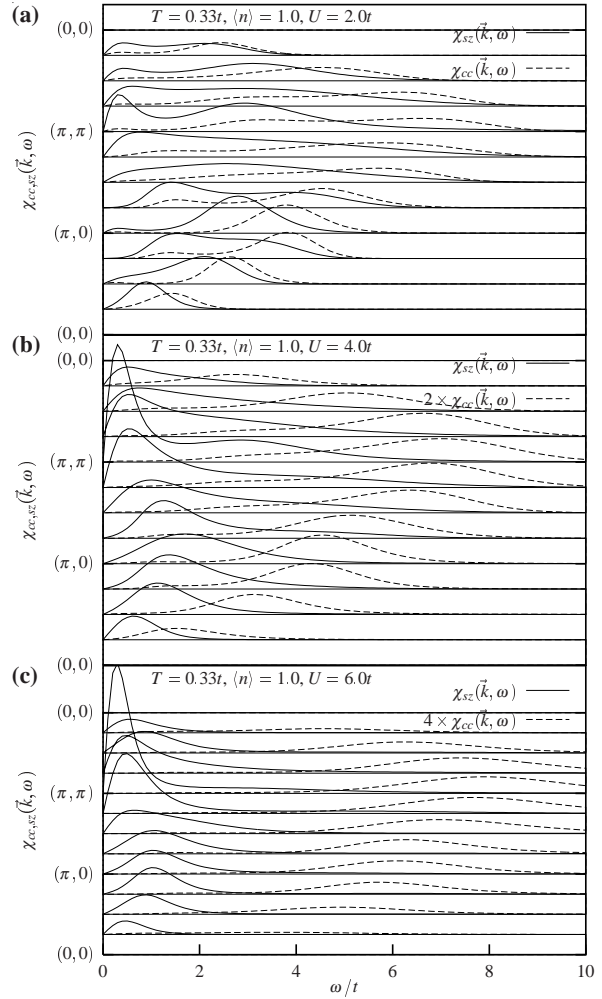,width=8.0cm,angle=0.0}
\vspace{0.0cm}
\narrowtext
\caption[]{Dynamical spin and density correlation function
for different values of $U/t$. Note that  the DFC is multiplied by
a factor of $2$ for $U/t=4$ and by a factor of $4$ for
$U/t=6$.}
\label{fig5} 
\end{figure}

 
\begin{multicols}{2}
Recently there has been renewed interest in the Mott-Hubbard
metal-insulator transition as a function of the interaction strength
in narrow-band systems\cite{jarrell,pruschke,kotlreview,gebhard}. 
At low temperatures the metal-insulator
transition is usually `masked' by ordering phenomena such as
antiferromagnetism or possibly orbital ordering. The insulating phase then 
is present even for infinitesimal interaction strength and can be explained 
in a simple one-particle picture in that ordering leads to a back-folding of 
the Brillouin zone and thus turns a half-filled band in the extended zone
into a completely filled one in the reduced Brillouin one. This is
no longer the case above the ordering temperature. In that sense
the high-temperature behavior is more interesting, because
it represents a pure correlation effect.\\
Motivated by these considerations we have performed a Quantum Monte Carlo 
(QMC) study of the  2D Hubbard model on the square lattice at elevated 
temperature, $T$$=$$0.33t$. While this temperature still is relatively 
small compared to the bandwidth $W$$=$$8t$ or `typical' values of $U$, it is 
relatively large on the scale of the exchange constant $J$$=$$4t^2/U$, so
that longer-ranged antiferromagnetic correlations essentially cease to exist.
The spin correlation length is $< 1.5$ lattice spacings 
for all $U$ under consideration (see Figure \ref{fig1})
and thus significantly smaller than the system size (at least $8\times8$).
This excludes `finite-size gaps' which typically occur\cite{vekic} when the
magnetic correlation length becomes comparable to the system size.
One might thus expect that the results obtained in this study are
quite representative for a correlated electron
system above the ordering temperature.
We present data for the single particle spectral function and the
dynamical spin and density correlation function. The data clearly show
a metal-insulator transition as a function of increasing $U$,
which sets in at the value $U$$=$$4t$, i.e. 50\% of the noninteracting
bandwidth. The Mott-Hubbard gap opens gradually as $U$ increases, and the
transition is accompanied by a rather discontinuous
change of the band structure. On the other hand,
we see no indication of a gradual
`band narrowing' or a vanishing of the quasiparticle 
weight throughout the 
metallic phase. Most spectacularly perhaps, the transition is accompanied 
by the formation of well-defined collective modes in the spin and charge
response functions.
\noindent 
To begin with, Figure \ref{fig2} shows the single particle spectral function
$A(\bbox{k},\omega)$ for small $U/t\le3$. For these small interaction 
strengths $A(\bbox{k},\omega)$ is almost completely
consistent with a noninteracting electron system. 
In particular, 
there is no distinguishable `correlation narrowing' of the bandwidth,
nor is there any appreciable reduction of the spectral weight in the
peaks near the Fermi energy. The main effect of
$U$ is an increased width of the peaks, which
usually is more pronounced at higher excitation energies, 
and thus can be understood as a kind of Landau damping. 
For the trivial case, $U$$=$$0$, the spectra give a feeling
for the `intrinsic MaxEnt broadening' due to inaccuracies of the MaxEnt
procedure. The bandwidth
between $(0,0)$ and $(\pi,\pi)$, however,  remains unaffected at $8t$.
To be more precise, we have computed the `average excitation energy',
i.e. the ratio of first and zeroth moment, for
the electron removal spectrum at $(0,0)$ and the electron addition 
spectrum at $(\pi,\pi)$. Taking the difference of these two
average energies to obtain
an `effective bandwidth' $\bar{W}$ gives the values
$\bar{W}$$=$$8.02$, $8.04$, $8.14$ and $8.34$ for $U/t=0,1,2,3$.
The width of the distribution of spectral weight thus even
{\em increases} slightly with $U$.
The probability for a doubly
occupied site, $\bar{d}$$=$$\langle n_{i\uparrow}n_{i\downarrow}\rangle$,
decreases quite strongly with $U/t$: we find $0.250,0.222, 0.194,$ and
$0.168$ for $U/t=0,1,2,3$. 
Using $\bar{d}$ to compute the Gutzwiller
renormalization factor for the bandwidth, $\gamma=8(1-2\bar{d})\bar{d}$,
we obtain the estimates $W_{Gutzwiller}$$=$$8.0, 7.90, 7.60$ and
$7.14$. The numerical data are quite obviously not consistent
with this.\\
The free electron form of $A(\bbox{k},\omega)$ changes appreciably
only as we reach $U$$=$$4t$ (Figure \ref{fig3}a). 
While the most intense `peaks' still can be
fitted roughly by a standard 
\noindent
cosine band with undiminished bandwidth $\ge 8t$, there now appears some
additional fine structure in the spectra. First, faint bands
with weak intensity form close to the Fermi
energy (they can be seen most clearly at
energies $\omega =\pm2t$ for $(0,0)$ and $(\pi,\pi)$),
and also a flat low-intensity band at $\omega$$=$$\pm 4t$ appears.
The topmost and lowermost parts of the original cosine band
now appear to be split, which increases the
total bandwidth to $\approx 10t$.
Secondly, and most spectacularly, the large low-energy peaks around
$(\pi,0)$ now show an unambiguous splitting. The peak at $(\pi/2,\pi/2)$ 
on the other hand is rather broad, which may indicate a tendency towards 
splitting, but it still has its maximum right at $E_F$.\\
Inspection of the further development of  $A(\bbox{k},\omega)$
for larger $U$ (Figures \ref{fig3}b,c)
shows that what actually happens at $U$$=$$4t$ is the formation
of a total of $4$ `bands'. There are two weak bands with a width
of $\approx 2t$ which form the first ionization/affinity states,
and two essentially dispersionless bands at higher energy.
Increasing $U$ makes this $4$-band structure more and more obvious
and increases the Mott-Hubbard gap. Apart from
a more or less rigid shift the dispersion of each of the
$4$ bands remains unaffected by the increase of $U$.
Let us stress that none of the $4$ bands show any
indication of antiferromagnetic symmetry, i.e. symmetry under the
exchange $\bbox{k}\rightarrow \bbox{k}+(\pi,\pi)$. This is 
(in addition to the short spin correlation length)
further evidence that the opening of the gap is
unrelated to antiferromagnetic broken symmetry.
Figure \ref{fig4} shows the spectra at $(\pi,0)$ and $(\pi/2,\pi/2)$
for different system sizes. The gap width
(defined e.g. as distance between the two maxima next
to $E_F$) is independent of system size, which demonstrates
that due to the short spin correlation length finite-size effects are 
negligible already at $L$$=$$8$. Independent of the system size, the gap
is slightly
\noindent 
smaller at $(\pi/2,\pi/2)$. The single particle
spectrum thus shows that different portions on the Fermi surface 
react differently to the increasing $U$. Interpreting the splitting
of the peaks at $(\pi,0)$ as the first signature of the
Mott-Hubbard gap, the data therefore suggest that the Mott-Hubbard
gap does not open uniformly over the whole Fermi surface but gradually
in different sections of the Fermi surface.
This is actually quite reminiscent of the `pseudo gap' phenomenology
in cuprate superconductors.
The initial opening of the Mott-Hubbard gap at $(\pi,0)$ suggests 
to draw a parallel to the `patch model' of Furukawa and 
Rice\cite{furukawarice}. These authors explained
the pseudogap in cuprate superconductors by enhanced scattering
between `patches' of momenta around van-Hove singularities (vHs) at 
$(\pi,0)$. In the model of Furukawa and Rice the kinetic energy was
augmented by a $2^{nd}$-nearest neighbor hopping integral $t'$, so as to
shift the vHs below the half-filled
Fermi surface. The vHs therefore is right at the Fermi energy only for
finite doping. In the absence of $t'$, as in the present case,
the vHs is right at $E_F$ for half-filling. Since the vHs is still at 
$(\pi,0)$, 
the theory of Furukawa and Rice may still be applicable and would
then nicely explain the initial opening of the gap at $(\pi,0)$.
\noindent 
We turn to the dynamical spin correlation function (SCF)
and density correlation function (DCF), which are shown in 
Figure \ref{fig5}. For the small value $U$$=$$2t$ SCF and DCF 
look nearly identical,
and show the broad continua expected for a nearly-free electron system.
At $U$$=$$4t$, on the other hand, the shape of SCF and DCF indicates
that despite the relatively free-electron like shape of $A(\bbox{k},\omega)$
the system is actually already very far from a free electron system: the SCF
shows only low energetic peaks, which roughly resemble the
spin wave dispersion of the insulator. 
The spectral weight, however, while being peaked at $(\pi,\pi)$,
is much more uniform along this `spin wave' branch than for the
insulator, consistent with the spin correlation length
of $<2$ lattice spacings. Despite the short correlation length
we thus have a quite well-defined collective spin excitation mode.
This mode is sharp and well-defined
even for the smallest wave vectors possible in our cluster, e.g. at
$(\pi/4,0)$. This is clear evidence that this mode is unrelated
to even short range antiferromagnetic correlations, because
in this case the
wavelength, $\lambda$$=$$8$, exceeds the spin correlation length by
more than a factor of $4$. The long wavelength would suggest that
this is actually a hydrodynamic mode - in this case, however,
it is quite remarkable that this excitation is apparently
`continuously connected' to the short wavelength excitations
at $(\pi,\pi)$.\\
At the transition, $U$$=$$4t$, the
density correlation function shows another spectacular feature, namely
the total absence of any low energy spectral weight at
or near the Fermi surface nesting vector $(\pi,\pi)$ - rather, there is
now a broad band of intense spectral weight whose center of gravity
can be fitted remarkably well by
a simple cosine-band of total width $8t$ and minimum at $(0,0)$. 
Again, 
\noindent 
this shows that the system is already very far from
any single-particle physics. Near $(\pi,\pi)$ the band in the DCF
has some `tails' towards lower energy, but
otherwise the relative sharpness of the band suggests that
this is another (and very different) collective excitation
as compared to the spin mode. As $U$ increases the band is shifted to
higher energy and its width decreases.
We note that the very different behavior of the spin and density correlation
function and even the cosine-band like shape of the DCF itself
are quite reminiscent of results for the {\em doped}
Mott-Hubbard insulator at higher $U$\cite{corr}.\\
Summarizing the data on the correlation functions it seems that
the hallmark of the transition to the insulating state is the
emerging of relatively well-defined collective modes in the
spin and density correlation function. The diffuse particle-hole continua,
which could be seen in the (nearly identical) spin and density
correlation functions at small $U$, give way to
rather sharp and `single particle like' modes
at larger $U$. It is tempting to speculate that
scattering of the electrons from these two different
collective modes may actually be responsible for the
$4$-band structure observed in the single-particle spectrum: 
scattering from the spin mode might produce the two
low energy bands and ultimately be responsible for the opening of the
Mott-Hubbard gap (which initially is rather a `d-wave pseudogap'
that opens only in certain parts of the Fermi surface), whereas
scattering from the charge-mode may produce the two high-energy bands.\\
To conclude, we turn to a comparison with the
$d\rightarrow \infty$ limit
or dynamical mean-field calculations\cite{metzner,kotl}.
Georges {\em at al.}\cite{kotlreview} have thereby developed a scenario
where the  metal insulator transition is preceded by a transfer of spectral 
weight from the region around the Fermi energy to precursors of the
Hubbard bands. In the spectral density, the metal-insulator transition 
occurs via the disappearance of a `Kondo resonance'-like
peak, which is in the center of a preformed gap -
the Mott-Hubbard gap thus
jumps discontinuously from zero to a value of order $U$.
Our data do not show much similarity with this scenario.
Neither is there any appreciable
decrease of the spectral weight in the peaks at the
Fermi energy, nor is there any indication of a preformed gap,
nor is there any discontinuity of the gap when it opens.\\
More recently, the scenario put forward by Georges {\em at al.} has
met some criticism\cite{logannozieres,kehrein,nohard}.
Kehrein\cite{kehrein} has recently argued that under the assumption
of Fermi liquid behavior in the metallic phase the scenario
of Georges  {\em at al.} is incorrect even in $d\rightarrow \infty$
One of the alternative scenarios put forward by Kehrein
was `semimetal-behavior', where at the transition there is finite
spectral weight in any neighborhood of the
Fermi surface - which is quite consistent with the pseudogap
behavior in our data. A numerical study by Noack and Gebhard\cite{nohard}
where the infinite dimension limit was simulated by
a random dispersion has also shown a continuous opeening of the
gap.\\
In summary, we have performed a QMC study of the paramagnetic
metal-insulator transition in the 2D Hubbard model. At 
moderately high temperature, $T$$=$$0.33$, antiferromagnetic
correlations are short ranged and
upon increasing $U$ our data show a metal-insulator transition.
It should be noted, that due to the finite temperature,
our data have little or no significance for a hypothetical paramagnetic
metal-insulator transition at zero temperature. Equally,
we cannot distinguish whether a gap is absent due to a too
small $U$, or due to a too high temperature. It has to be kept in mind, 
however, that the hypothetical zero-temperature phase transition in most
cases is superseded by ordering phenomena anyway,
so that the finite temperature transition observed here
is probably the more physically relevant one.\\
We then found that the single particle spectrum in
the metallic phase looks `unusually unusual', in that despite
a significant reduction of charge fluctuations, the
only effect of the increasing interaction seems to be
an increased broadening of the peaks. There is no indication
for any `correlation narrowing' of the bandwidth, or for a decrease of
the spectral weight at the Fermi energy. Since
we are using a finite mesh of $\bbox{k}$-points
we cannot rule out an increase of the band mass only
in a very small neighborhood of $E_F$. A global band narrowing, however,
is inconsistent with our data.\\
The Mott-Hubbard gap opens continuously, and in fact
nonuniformly on the Fermi surface.
In the very first stage of gap formation the gap 
seems to open at $(\pi,0)$, which may
suggest a connection with the `pseudogap' in cuprate superconductors, as
described by the `patch model' of Furukawa and Rice.
There is no preformed gap or any discontinuity in the Hubbard gap.
At the transition we observe the formation of surprisingly sharp collective 
modes in the spin and density response. These appear to be the hallmark of 
the insulating state, and initially may even drive the gap formation.\\
We thank F. Gebhard for helpful comments.
This work was supported by DFN Contract No. TK 598-VA/D03, by BMBF
(05SB8WWA1),
computations were performed at HLRS Stuttgart, LRZ M\"uchen and HLRZ J\"ulich.
 
\end{multicols}
\end{document}